\documentclass[sigplan,screen]{acmart}
\AtBeginDocument{%
  }

\setcopyright{acmlicensed}
\copyrightyear{2018}
\acmYear{2018}
\acmDOI{XXXXXXX.XXXXXXX}
\acmConference[Conference acronym 'XX]{Make sure to enter the correct
  conference title from your rights confirmation email}{June 03--05,
  2018}{Woodstock, NY}
\newcommand{\flashx}{Flash-X}
\newcommand{\codescribe}{CodeScribe}
\newcommand{\mcfm}{MCFM}
\usepackage{graphicx}
\usepackage{subcaption} 
\usepackage{caption}
\newcommand\codetext[1]{{\small\texttt{#1}}}
\usepackage{algorithm}
\usepackage{algpseudocode}

\begin{document}

\title{Adding New Capability in Existing Scientific Application with LLM Assistance}

\author{Anshu Dubey}
\email{adubey@anl.gov}
\orcid{0000-0003-3299-7426}
\affiliation{%
 \institution{Argonne National Laboratory}
 \city{Lemont}
 \state{IL}
 \country{USA}}

\author{Akash Dhruv}
\email{adhruv@anl.gov}
\orcid{0000-0003-4997-321X}
\affiliation{%
  \institution{Argonne National Laboratory}
  \city{Lemont}
  \state{IL}
  \country{USA}}

\renewcommand{\shortauthors}{Trovato et al.}

\begin{abstract}
With the emergence and rapid evolution of large language models (LLM), automating coding tasks has become an important research topic. Many efforts are underway and literature abounds about the efficacy of models and their ability to generate code. A less explored aspect of code generation is for new algorithms, where the training data-set would not have included any previous example of similar code. In this paper we propose a new methodology for writing code from scratch for a new algorithm using LLM assistance, and describe enhancement of a previously developed code-translation tool, Code-Scribe, for new code generation.
\end{abstract}




\maketitle


\section{Introduction}
Advent of increasingly more sophisticated large language models (LLMs) {e.g. \cite{chatgpt,Codellama2024,copilot,huggingface,jiang2023mistral7b} has created immense interest in exploring how and how much can they aid in software development. Many projects are exploring various aspects of code generation and speculations are rife about the extent to which software development can be automated \cite{Dearing2024,slack,godoy2024,teranishi2025,zi2025}. A less explored, and less obvious aspect of code generation is when the software to be developed is about a new algorithm or a new concept. The challenge arises out of making the specification precise enough for the model to generate desired code. Prompts are supplied in natural languages, which are imprecise by nature, i.e. same sentence can be interpreted in more than one way depending upon the context. Also, humans internalize knowledge which is often skipped in writing specification. A typical outcome of imprecise specification is model hallucination, from which it can be difficult, if not impossible, to recover. 

In this paper we introduce a new methodology for prompt engineering that can significantly reduce model hallucination. The methodology is based on interacting with the model iteratively on specifications rather than code, repeatedly asking the model to articulate what it understood to be the {\em ask} from it, and correcting it as necessary. Since this interaction proceeds in natural language on both ends, it becomes much easier to catch and correct incomplete or erroneous specifications. We used and enhanced \codescribe, an LLM-assisted tool \cite{Dhruv2025} that was originally developed for translating large legacy code from Fortran to C++, as the mechanism for interacting with the models because it enables chat-completion though the use of TOML files which are easy to read, modify, and maintain. We describe the enhancements and methodology in the context of a new technique of handling interaction between cartesian mesh and Lagrangian particles that we are developing for \flashx \cite{DUBEY2022101168}, a multiphysics software system used for simulations by multiple domains.  

The remainder of the paper is organized as follows -- section \ref{codescribe} gives a brief description of the main features of \codescribe~that make it an attractive tool to use in general coding related tasks. Section \ref{algorithm} describes the new communication algorithm that motivated the work described here, the methodology we developed for code generation, and enhancements to \codescribe~to enable its use for code generation. In section \ref{experiments} we describe our experiences in using different models in various modes of interaction, and in section \ref{conclusions} we summarize our observations and conclusions.
\section{\codescribe}
\label{codescribe}
The tool was designed in response to the challenge of converting legacy fortran code \mcfm \cite{Campbell_2019,Campbell_2011}, used for modeling interactions in Large Hadron Collider (LHC), into C++. Therefore, a large part of the tool is focused on handling the organization of the source code. It does so by indexing subroutines, modules, and functions across various files to understand the relationships between components in the code. Patterns are identified in the original Fortran code and are used as guide for creation of chat templates. The templates serve as references in translating files with similar source code patterns. It then creates a draft of the C++ code for a given Fortran source file. It employs pattern recognition to replace variable declarations, include appropriate header files, and identify the use of external modules and subroutines. This information is encoded in the draft to help LLMs better understand the code context. Using the chat template as a seed prompt, \codescribe~appends the source and draft files and triggers the LLM to complete the code conversion. The generated results are extracted, reviewed, compiled, and tested for correctness. Any errors found during compilation are addressed manually by the developer or sent for regeneration by updating the seed prompt. To aid in translation the tool has a command-line interface with four commands: (1) {\em Index}, (2) {\em Inspect}, (3) {\em Draft}, and (4){\em Translate}. Detailed documentation for the usage of these commands can be found in the \codescribe~repository \cite{akash_dhruv_2024_13879406}; and \cite{Dhruv2025}

A new algorithm that does not already have any existing source does not need the first three commands of \codescribe. The {\em Translate} command, being the primary AI-powered portion of the tool, provides a template for adding a new command {\em Generate}. In {\em Translate} command a seed prompt from the chat template is appended with code from the source and draft files and sent to the AI for chat completion. The source code from the Fortran file is supplied within the elements \codetext{<source>...</source>}, and the draft code is enclosed within \codetext{<draft>...</draft>}. The LLM is instructed to produce output for the C++ source and the corresponding Fortran-C interface, enclosing them within \codetext{<csource>...</csource>} and \codetext{<fsource>...</fsource>}. The {\em Generate} command, having no existing code, just uses the prompt without any code being appended. In every other respects it executes the same way as the {\em Translate} command. Interaction with the LLM is achieved through direct API endpoints or the Transformers library. 

\section{Algorithm}
\label{algorithm}
Many astrophysics simulations use N-body methods combined with a mesh where mesh carries density as a variable, and Lagrangian particles represent mass. Before calculation of gravitation potential by the mesh using density, the particles deposit density onto the mesh from the mass they are carrying. We use a simple cloud-in-cell (CIC) \cite{BIRDSALL1997} method where cells that receive the density from a particle include one where the particle is located and its immediate nearest neighbors along each axis. The criterion for which side neighbor gets the density is:
\begin{eqnarray}\nonumber
    particle\_pos(ax)&<=& cell(ax) -> neigh=cell(ax-1) \\ \nonumber
    particle\_pos(ax) &>& cell(ax) -> neigh=cell(ax+1) \\ \nonumber
    where~ax &\in &\{i, j, k\} \nonumber
\end{eqnarray}
An example of deposition with CIC method on a 2D mesh is shown Figure \ref{fig:cic}. The left side of the figure shows a particle that is in the interior of a domain where the darker colored cells receive a fraction of the mass as density such that the total deposited on all the cells adds up to 1. The right side of Figure \ref{fig:cic} shows a situation where the particle is located on a cell adjacent to a physical boundary. Special handling is needed to account for boundary conditions (BCs) in such a situation. For most BCs both the cells inside the domain will get the same fraction as they would have if they were interior cells and the remaining mass will be lost. Periodic BCs are different, they assume that the domain wraps around. Here, cells at the opposite end of the domain are considered adjacent, and they will get the appropriate fraction of the mass as shown in the right side of the figure. When the domain is divided into blocks, the adjacent cells in the neighboring block will have the deposition.

\begin{figure}[h] 
  \centering
    \includegraphics[width=\linewidth]{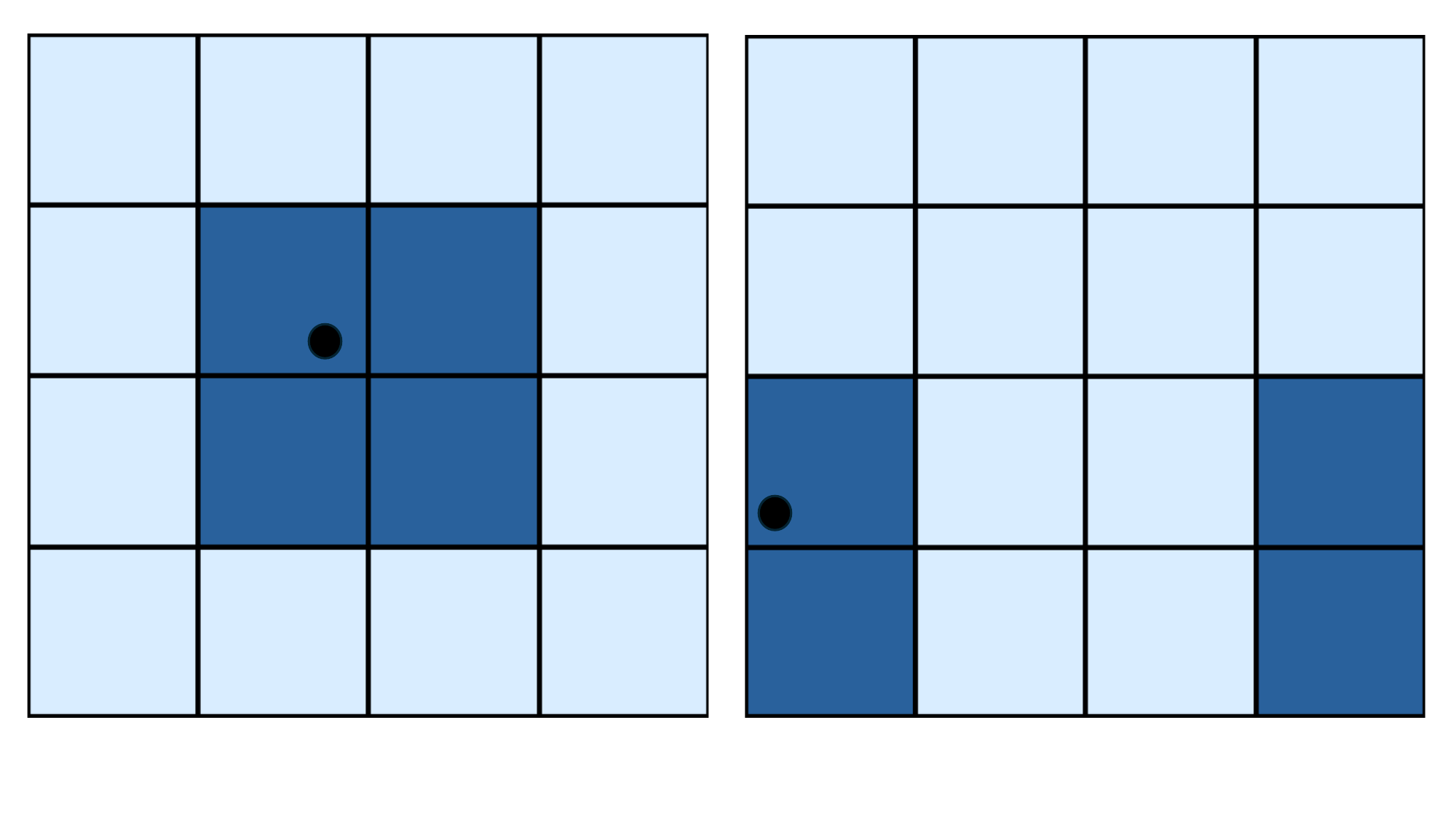}
    \caption{Deposition from a particle onto the mesh. The darker cells are where the deposition occurs. The left panel shows a situation where the particle lies on an interior cell, while the right panel shows a particle that lies on a cell at the boundary, and periodic boundary conditions are assumed.}
  \label{fig:cic}
\end{figure}

\subsection{Parallel Alogithm Overview}

\begin{figure}[h] 
  \centering
 \includegraphics[width=\linewidth]{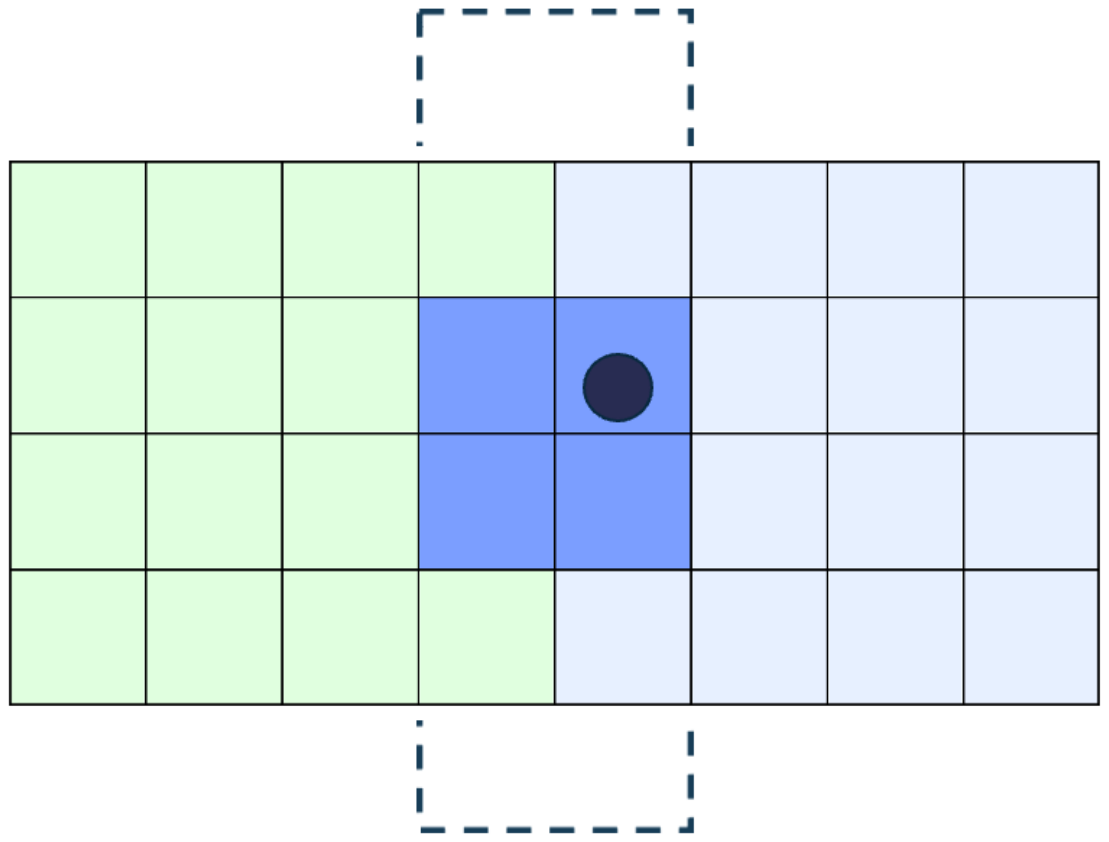}    
  \caption{Deposition on the edge of the block where halo cells get some of the mass from the particle as density.}
  \label{fig:pic3}
\end{figure}

\flashx~uses adaptive mesh refinement (AMR) \cite{BERGER1984} where resolution varies in the domain depending upon the demands of the flow. Where the flow is smooth resolution is low, where there are steep gradients (e.g. in a shock) resolution is high. This is achieved by decomposing the domain into blocks where each block has identically sized cells, but the size of cells differs in different blocks.  Explicit solvers, often used in astrophysical flows, have their blocks surround by a halo of ghost-cells that are filled with values from the neighboring blocks so that stencil computations can occur locally. In an earlier version of \flashx, FLASH \cite{dubey2009extensible,dubey2014evolution}, same halo cells that are used for stencil updates were used to deposit values from a particle on a block boundary as shown in Figure \ref{fig:pic3}. After the deposition a reverse halo cell filling operation ensured that the deposited values reach the correct destination cells. As the solution evolves, particles themselves move within the domain, and sometimes migrate between blocks. A communication algorithm moves the particles to their correct destination block when such a migration occurs \cite{dubey2012imposing,DUBEY2011101}. Thus two distinct communication steps described in algorithm \ref{alg1} are involved in simulations that include massive particles. 
\begin{algorithm}
\caption{Particle Advance, Migration, Deposition, and Reverse Ghost Fill}
\label{alg1}
\begin{algorithmic}[1]
\Require Particle set $\mathcal{P}$, timestep $\Delta t$, block map $B(\cdot)$
\State $\mathcal{L} \gets \varnothing$ \Comment{local container}
\State $\mathcal{D} \gets \varnothing$ \Comment{destination buffer}
\ForAll{$p \in \mathcal{P}$}
  \State \textbf{Advance} $p$ by $\Delta t$
  \If{$p$ moved out of its block}
    \State Update $B(p)$ \Comment{recompute block ID}
    \State $\mathcal{D} \gets \mathcal{D} \cup \{p\}$
  \Else
    \State $\mathcal{L} \gets \mathcal{L} \cup \{p\}$
  \EndIf
\EndFor
\State \Call{MoveParticles}{$\mathcal{D}$} \Comment{communication/migration}
\State $\mathcal{R} \gets$ received particles
\State $\mathcal{L} \gets \mathcal{L} \cup \mathcal{R}$
\ForAll{$p \in \mathcal{L}$}
  \State Deposit density from $p$ into overlapping cells
\EndFor
\State \Call{ReverseGhostCellFill}{}
\end{algorithmic}
\end{algorithm}

We propose a new method that eliminates the need for the reverse ghost-cell fill  step which can be very costly in AMR because of irregular communications. This can be done by creating virtual copies of a particle that is on a cell adjacent to one or more block boundaries. During the step to move the particles to their new destination all the virtual copies can be appended to the list of particles being moved to a remote processor. The virtual particles then deposit directly into their destination cells, and are destroyed once the deposition is done. This may add some cost to the operation of moving particles because the amount of data being sent is larger, however, the number of particles crossing processor boundaries is usually very modest, so we expect latency costs to dominate. Similar argument applies to extra space being used by the virtual particles.
We describe the algorithm using a simplified use-case  where the domain is decomposed into blocks of equal size and identical resolution.  Correct handling of deposition from the virtual particles is more complicated with AMR, but is out of scope for this work where the focus is on LLM assistance in code writing. Steps in the new proposed algorithm are described in algorithm \ref{alg2}. In \flashx~ particles are stored in a two-dimension array where the first dimension contains various attributes associated with the particle. Here, we are only concerned with the coordinates of a particle, and the block that the particle is associated with. To handle virtual particles we add another attribute to the particle which is set to $1$ if the particle is virtual, and $0$ if it is real. We need this attribute to be able to destroy virtual particles when they are done with deposition.

\begin{algorithm}
\caption{Particle Advance, Virtual Particle Generation, Migration, and Deposition}
\label{alg2}
\begin{algorithmic}[1]
\Require Particle set $\mathcal{P}$, timestep $\Delta t$, block map $B(\cdot)$, boundary distance threshold $\tfrac{1}{2}\Delta$
\State $\mathcal{L} \gets \varnothing$ \Comment{local container}
\State $\mathcal{D} \gets \varnothing$ \Comment{destination buffer}

\ForAll{$p \in \mathcal{P}$}
  \State \textbf{Advance} $p$ by $\Delta t$
  \If{$p$ is within $\tfrac{1}{2}\Delta$ of any block boundary}
    \State Generate virtual particles $\mathcal{V}_p$ around $p$
  \Else
    \State $\mathcal{V}_p \gets \{\,\}$ \Comment{no virtuals}
  \EndIf
  \State $\mathcal{S}_p \gets \{p\} \cup \mathcal{V}_p$ \Comment{real + virtual particles}
  \ForAll{$q \in \mathcal{S}_p$}
    \State Find new block $B(q)$ based on position of $q$
    \If{$B(q)$ equals current block of $p$}
      \State Add $q$ to $\mathcal{L}$
    \Else
      \State Add $q$ to $\mathcal{D}$
    \EndIf
  \EndFor
\EndFor

\State \Call{MoveParticles}{$\mathcal{D}$} \Comment{perform communication/migration}
\State $\mathcal{R} \gets$ received particles
\State $\mathcal{L} \gets \mathcal{L} \cup \mathcal{R}$

\ForAll{$q \in \mathcal{L}$}
  \State Deposit density from $q$ into overlapping cells
  \If{$q$ is a virtual particle}
    \State Destroy $q$
  \EndIf
\EndFor

\end{algorithmic}
\end{algorithm}

Next we describe the methodology for developing code with assistance from LLMs. We use LLM assistance for the overall code development up to the point where we have all the virtual particles created and sorted. This is because the remainder of the steps (for moving particles and deposition of density) remain largely unchanged from the original, already existing code for a uniform mesh. The only additional step required in the new method is the destruction of virtual particles to avoid memory bloat. In all our code generation the  assumption is that at the beginning of a timestep we have only real particles and they all have correct block ID in their attributes. With time integration some particles may move within the block and some may move out of the block. The simulations of interest apply a  constraint to time-steps where a particle may not move more than one cell in one timestep. Therefore, when we begin to process particles some may be outside the block, but they are likely to be close enough to the block boundary that one of their mirrors may lie within the block. 

\subsection{Methodology}
\label{methodology}
Using assistance from LLMs in coding tasks requires having robust verification in place because  correct code generation is not guaranteed. Therefore, we followed test-driven development (TDD) principles for the entire project. Our testing infrastructure started with a mesh shown in Figure \ref{fig:mesh} where the mesh is divided into 16 blocks of equal size and the blocks are assigned an integer ID (in the center of the block). The figure also shows the coordinates bounding each block (referred to as block-bounds from here on).
\begin{figure}[h] 
  \centering
 \includegraphics[width=\linewidth]{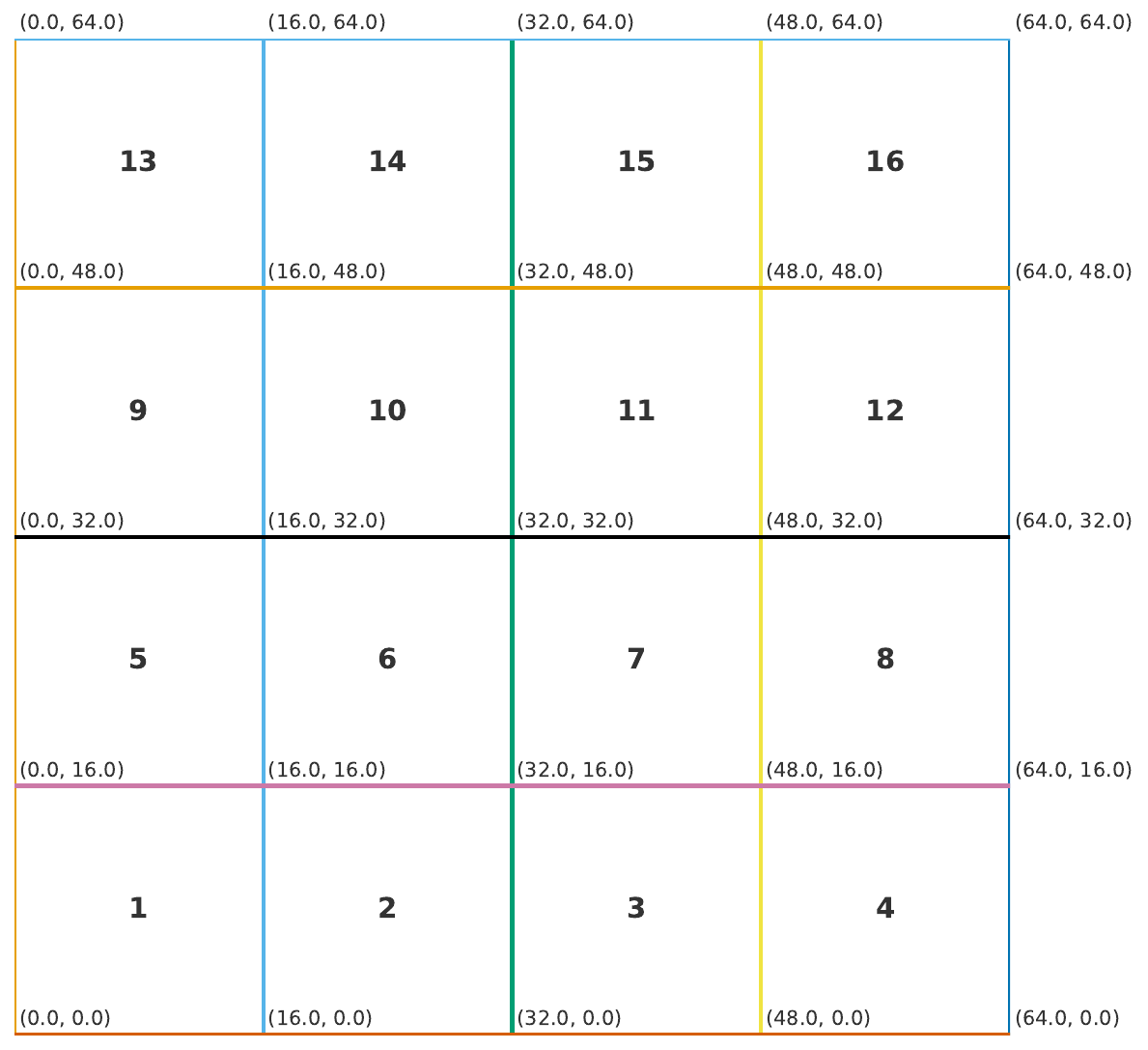}    
  \caption{Mesh used for testing the code as we develop it.}
  \label{fig:mesh}
\end{figure}
We next generated some helper functions that were necessary for testing. These included a function to get the lower and upper bounds of the block given its ID, and also to get the ID of the overlapping block from the coordinates of a particle. All of these functions already exist in \flashx, so we could work with highly simplified versions for testing. As a result these functions required very simple prompts and the generated code was largely correct. In some instances there were compilation errors, many of which stemmed from incomplete specification, and were easily fixed. Next we broke down the overall coding effort into tasks that could be verified individually. The first task is to generate mirroring positions for the virtual particles. In a 2D mesh, a particle that is close to the block-bounds along only one axis will have one mirror, while particle that is close to a corner will have three mirrors, one along each axis and one diagonally across. Following the same logic in a 3D mesh a particle can have 1, 3, or 7 mirrors depending upon whether it close to a face, an edge or a corner. 

In a real simulation, (BCs) at the physical boundaries of domain also affect the number of mirrors, therefore, the next task is to enable handling of BCs. Note that different axes can have different BCs, therefore each axis much be treated independently. If the BC is periodic along any axis then there is a mirror along the axis at the opposite end of the domain. If all axes have periodic BCs then the same logic of face, edge and corners applies as in the interior blocks for the number of mirrors, the only difference is in the location of the mirroring positions. If the particle is just outside the domain in an axis with outflow BC the particle is effectively lost and no virtual particles are needed. However, if the particle is still within the domain, but near the boundary then the real particle is still in action but it will not have any virtual particles along the axis where the BC is outflow. For an axis with reflective BC the real particle goes back into the domain and does not need virtual particles. Other, more complex BCs may arise, for the purpose of this code development we accounted only for the three BCs mentioned above.  

The final task is to assign correct attributes to the particle so that the deposition occurs correctly. Note that even though the mirrors are created at positions different from that of the original particle, for correct deposition each one of the virtual particles needs to have the coordinates of the original real particle. The mirrored positions are temporary, and are used only to determine the block ID of the neighboring block where the particle needs to deposit. Normally a particle will be associated with a block whose bounds include the coordinates of the particle but virtual particles violate this requirement. They carry ID of the block that would have received the cells in reverse ghost-cell fill step. We used several different approaches for developing prompts for these tasks with mixed success for a while. Our final approach of extending and leveraging \codescribe~has been the most successful and has worked on various models we have tried so far. Also, during the course of the development we changed the design of the code several times and abandoned partial code when a better design occurred to us -- something that is not possible when one is writing code manually. In the next section we sketch out design iterations and various approaches for interacting with the LLMs

\section{Experiments }
\label{experiments}
 In this work we have worked with 3 models: ChapGPT \cite{chatgpt}(we started with 4, and then 5 became available halfway through the project), and Codellama \cite{Codellama2024} and Kimi (another model hosted internally) available through web-interface at RIKEN R-CCS in Japan. Both the models at RIKEN are tuned for coding related tasks. 

In the first iteration of code design we opted to determine whether the particle is outside the block or inside, and if inside, where is it located within the associate block. We used a variable $edge(NDIM)$ where NDIM is the mesh dimension. For each axis $ax$ $edge(ax)$ would get a value OUTSIDE/INTERIOR/LOW/HIGH depending on whether particle is outside the box, inside the box but away from both ends of the box, or within $\delta/2$ distance of either lower or upper end of the box along $ax$. We would then determine how many axes had a value other than INTERIOR to determine the number of mirrors to be created. Finally it would go ahead and create mirrors. At this point BCs were not being considered at all. We found that degree of detail in the prompt to be an overspecification. The prompt is unnecessarily long, it prevents the models from being able to infer the intent of the prompt at all, and any imperfection in the specifications leads to hallucinations. Therefore in the next iteration we left the model to decide how it would determine which particle should have mirrors. Even though this did not generate correct code immediately, it was much easier to explain to the model where the generated code was incorrect, and after a few iterations ChatGPT was able to generate correct code, but we did not succeed with Kimi. In these iterations we did not use codellama.

The next step in the process, where we introduced handling of BCs the interaction became more difficult. We considered two approaches. One was to apply boundary conditions before creating mirrors, and the second was to create mirrors and then remove and/or move those that would be eliminated by the BCs. The first approach was attractive because with outflow and reflective BCs virtual particles are eliminated along the corresponding axes. In particular with outflow BC, with the particle leaving the domain no further processing is needed. However, this approach failed with both models. The specifications became too convoluted and we never succeeded in getting them exactly right. The second approach was more successful because it cleanly separated the logic of the two operations, namely creating mirrors and applying BCs. Several branches that were necessary in handling BCs first are unnecessary when they only result in pruning already filled data structures. 

The real breakthrough came with the insight that once the model has generated a code, it is extremely difficult to convey the information about where and how the code is incorrect. Most of the time when the initially generated code contained an error either because of fault in the specification, or fault in the interpretation by the model, we would have to reset the process, modify the prompt offline and start a completely new chat. Even then the context would not be completely wiped clean, and one could see the model inject some of the information from the earlier interactions. Our final and most successful attempt at getting correct code was to change the entire mode of interaction. Instead of asking the model to write the code we asked the models to explain what they had understood about the problem being solved without writing any code. The responses came back in natural language interspersed with mathematical notations where appropriate. In this mode it was much easier to iterate over the specifications until we could be confident that they were precise enough. Here again, ChatGPT far outperformed the other models. Below we show excerpts from the prompt and the specification generation by ChatGPT. The deleted portions are largely those that specified constants and data structures that could be assumed, or the description of logic that we have already explained earlier.

\begin{verbatim}
The problem that I want to solve is related to 
particle-in-cell where the mesh in the domain 
is cartesian, and is divided into boxes. 
Particles carry physical quantity and deposit 
it onto the mesh. The cells that they deposit 
on are -- the cell on which the particle is 
located, and its neighboring cells. 
Which neighboring cells (left or right along 
any axis) depends upon which edge the particle 
is closer to. This routine is designed 
to handle the situation when a particle is less 
than 0.5*delta away ...(deleted text)....
Mirrors images of the particle are 
to be created in all relevant directions.
.....(deleted text)....
If any of the faces of the box are on the physical 
boundaries then the handling of mirrors changes as 
....(deleted text)
I would like you to explain how would you organize 
and write this code.
\end{verbatim}
The specifications originally generated by the model also assumed degenerate cases where the domain could be extremely thin and mirrors may be needed along both sides, so it came back with the maximum number being $8$ in 2D and $27$ in 3D. This is the same number of neighbors that a block can potentially have in 2D and 3D respectively. Figure \ref{fig:snap1} shows a snap shot of the next prompt and the response.
\begin{figure}[h] 
  \centering
 \includegraphics[width=\linewidth]{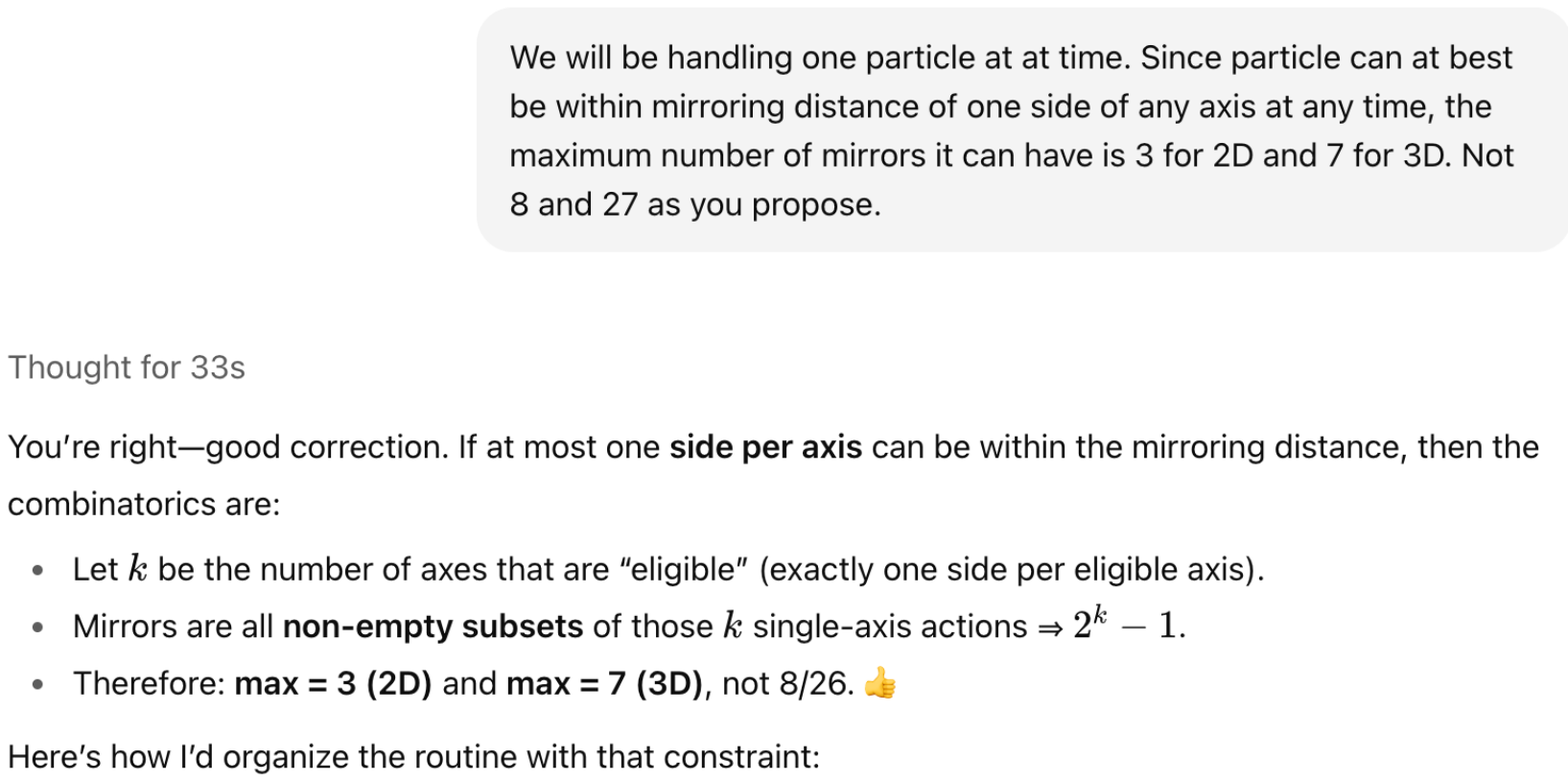}    
  \caption{Snapshot of interaction to correct specifications as understood by the model}
  \label{fig:snap1}
\end{figure}
After a few more iterations in correcting the specifications similar to one shown in Figure \ref{fig:snap1} the model was able to generate correct code. We also asked for a test program that would exercise the code. In the test the user would be prompted for the coordinates of a point and the test would print the mirrors generated. It was left to the user to inspect the results and verify them. The mesh shown in Figure \ref{fig:mesh} was the reference for verification. We saved the outputs generated by running the test to be used as baseline for experimentation with the other models. 

We did not find this level of reasoning sophistication with Kimi or Codellama. The prompt originally provided to the other models was identical to that provided to ChatGPT. With both of those models we had to abandon getting the specifications out of the model after several iterations.  Here, we finally succeeded by breaking down the problem into two tasks, where the logic of each task was simpler to explain. This was the approach of first creating all the mirrors and then removing ones that are eliminated by the BCs that we described earlier. Here, we used the following two prompts. The deleted text in the prompt is where it describes the logic for creating mirrors. With these prompts other models also generated correct code, and more importantly, the code is more compact than the one generated by ChatGPT.  

\begin{verbatim}
Prompt 1:
I want to write a routine that creates mirror 
images for a coordinate ....(deleted text) ....
For example, in a 3D mesh if  
blockbounds are  <0,0,0><10,10,10>, and 
cell size is <2,2,2> then the
following given points should return the 
mirrors as listed here.
<1,3,3> has only one mirror <-1,3,3>
<9,3,3> has only one mirror <11,3,3>
<3,1,3> has only one mirror <3,-1,3?
<3,3,9> has only one mirror <3,3,11>
<1,1,3> has three mirrors <-1,1,3>, 
<1,-1,3> and <-1,-1,3>
<3,1,9> has three mirrors <3,-1,9>,
<3,1,11> and < 3, -1, 11>
<1,1,1> has seven mirrors <-1,1,1>, 
<1,-1,1>, <1,1,-1><-1,-1,1>,
<1,-1,-1><-1,1,-1> and <-1,-1,-1>
<-1,9,1> has seven mirrors <1,9,1>, 
<-1,11,1>, <-1,9,-1><1,11,1>,
<-1,11,-1><1,9,-1> and <y1,11,-1>

Prompt 2: 
Next I want a routine to handle boundary conditions 
for each particle individually. Check if the 
position of the particle is outside the global 
domain contained and apply boundary conditions/
If the Boundary condition is periodic move the 
particle to the opposite face as one normally 
does for boundary condition. If Boundary condition 
is outflow, nothing is to be done, the particle is
lost. If the boundary condition is reflective 
then if  VP is false the particle is reflected 
back into the domain, and if VP is true then the
particle is lost. 
\end{verbatim}

In the final stage, we considered adding a new feature to \codescribe~that would leverage its chat completion capability for code generation. After adding the command {\em Generate}, we built the toml file for chat completion by appending both the user query and the model response to the file for every query–response pair. We followed the earlier methodology of prompting the model to generate a specification rather than code first. The toml file was iteratively edited whenever a misinterpretation in the "user" prompt caused incorrect generation.

The outcome of this exercise is a toml file for chat completion that, when used with \codescribe, produces correct and functional code across all three models tested. Interestingly, while the generated codes differ structurally, they all compile and behave correctly. 

Building on the mechanisms implemented in the {\em Translate} command, we extended \codescribe~to insert newly generated source code within \codetext{<source>...</source>} elements. This enables the system to automatically extract the generated code blocks and directly integrate them into Flash-X source files, eliminating manual intervention. The same metadata-aware parsing logic used for identifying \codetext{<csource>} and \codetext{<fsource>} elements is now reused for these \codetext{<source>} sections, ensuring that \codescribe~can seamlessly add, replace, or update code segments within the Flash-X codebase.

\section{Observations and Conclusions}
In this paper we have described a methodology for writing new code using LLMs. The paper is not about experimentation with different LLMs as its main research objective, instead, it is about finding a reliable way to generate code for new algorithms. In our use case, it is very likely that the models have seen similar concepts and/or code snippets in their training because none of involved concepts are by themselves very difficult or new. However, the way in which these concepts are used and combined in devising a new algorithm is new. Therefore, the challenge for this work was in getting the specifications right. As long as we interacted with the models by asking for code generation right away we had very limited success with getting correct code for our core problem. We did get correct code for our entire testing system without much difficulty because they are all fairly standard functions such as mesh generation, identifying overlap between a given coordinate and bounds of a section of the mesh etc. We have two primary insights from this exercise.
\begin{itemize}
    \item It is more productive to get the model to generate specifications by giving the description of the problem to be solved. One can iterate over the specifications before generating code. For both, the human-in-the-loop and the model, it is much easier to converge on common understanding of the task through modification of the specifications rather than code.
    \item Chat completions are very powerful in guiding the models because they preserve context. Having our chat completion in a toml file is even better because it permits correction to the context by inspecting and modifying the interaction offline. 
\end{itemize}

We find that use of LLMs to generate code through the use of tools like \codescribe~can improve both developer productivity and code quality. In our experimentation we found that every iteration of design shift resulted in cleaner and more concise specifications. Because actual writing of code itself does not take up much of developer time, it is easier to abandon code and start again. Also, because code generation by LLMs does not give any guarantee of correctness, software engineering practices such as rigorous verification and modular design become indispensable, resulting in better quality code. Additionally, one can ask the models to include comments in the code, which most developers tend not to include by default. Code generated by the LLM also includes checks for error conditions, another feature of good coding practices often ignored by scientific software developers. A very useful side-effect of using \codescribe~for code development is that with the chat preserved in toml file and inline documentation in the generated code, one has complete documentation for the code which is otherwise nearly impossible to have.

\label{conclusions}
\section*{Acknowledgments}

The submitted manuscript has been created by UChicago Argonne, LLC,
Operator of Argonne National Laboratory (“Argonne”). Argonne, a
U.S. Department of Energy Office of Science laboratory, is operated
under Contract No. DE-AC02-06CH11357. The U.S. Government retains for
itself, and others acting on its behalf, a paid-up nonexclusive,
irrevocable worldwide license in said article to reproduce, prepare
derivative works, distribute copies to the public, and perform
publicly and display publicly, by or on behalf of the Government.  The
Department of Energy will provide public access to these results of
federally sponsored research in accordance with the DOE Public Access
Plan. http://energy.gov/downloads/doe-public-access-plan. 
This work was funded by the Scientific Discovery through Advanced
Computing (SciDAC) program via the  Office of Nuclear Physics and
Office of Advanced Scientific Computing Research in the Office of
Science at the U.S.\ Department of Energy. Code generation was done using models from OpenAI and two Models internally housed at RIKEN R-CCS, Codellama and Kimi.

\bibliographystyle{ACM-Reference-Format}
\bibliography{biblio}


\begin{thebibliography}{21}


\ifx \showCODEN    \undefined \def \showCODEN     #1{\unskip}     \fi
\ifx \showISBNx    \undefined \def \showISBNx     #1{\unskip}     \fi
\ifx \showISBNxiii \undefined \def \showISBNxiii  #1{\unskip}     \fi
\ifx \showISSN     \undefined \def \showISSN      #1{\unskip}     \fi
\ifx \showLCCN     \undefined \def \showLCCN      #1{\unskip}     \fi
\ifx \shownote     \undefined \def \shownote      #1{#1}          \fi
\ifx \showarticletitle \undefined \def \showarticletitle #1{#1}   \fi
\ifx \showURL      \undefined \def \showURL       {\relax}        \fi
\providecommand\bibfield[2]{#2}
\providecommand\bibinfo[2]{#2}
\providecommand\natexlab[1]{#1}
\providecommand\showeprint[2][]{arXiv:#2}

\bibitem[{Albert Q., et al.}(2023)]%
        {jiang2023mistral7b}
\bibfield{author}{\bibinfo{person}{{Albert Q., et al.}}}
  \bibinfo{year}{2023}\natexlab{}.
\newblock \bibinfo{title}{Mistral 7B}.
\newblock
\showeprint[arxiv]{2310.06825}~[cs.CL]
\urldef\tempurl%
\url{https://arxiv.org/abs/2310.06825}
\showURL{%
\tempurl}


\bibitem[{Anshu Dubey, et al.}(2022)]%
        {DUBEY2022101168}
\bibfield{author}{\bibinfo{person}{{Anshu Dubey, et al.}}}
  \bibinfo{year}{2022}\natexlab{}.
\newblock \showarticletitle{Flash-X: A multiphysics simulation software
  instrument}.
\newblock \bibinfo{journal}{\emph{SoftwareX}}  \bibinfo{volume}{19}
  (\bibinfo{year}{2022}), \bibinfo{pages}{101168}.
\newblock
\showISSN{2352-7110}
\href{https://doi.org/10.1016/j.softx.2022.101168}{doi:\nolinkurl{10.1016/j.softx.2022.101168}}


\bibitem[{Baptiste Rozière, et al.}(2024)]%
        {Codellama2024}
\bibfield{author}{\bibinfo{person}{{Baptiste Rozière, et al.}}}
  \bibinfo{year}{2024}\natexlab{}.
\newblock \bibinfo{title}{Code Llama: Open Foundation Models for Code}.
\newblock
\showeprint[arxiv]{2308.12950}~[cs.CL]
\urldef\tempurl%
\url{https://arxiv.org/abs/2308.12950}
\showURL{%
\tempurl}


\bibitem[Berger and Oliger(1984)]%
        {BERGER1984}
\bibfield{author}{\bibinfo{person}{Marsha~J Berger} {and}
  \bibinfo{person}{Joseph Oliger}.} \bibinfo{year}{1984}\natexlab{}.
\newblock \showarticletitle{Adaptive mesh refinement for hyperbolic partial
  differential equations}.
\newblock \bibinfo{journal}{\emph{J. Comput. Phys.}} \bibinfo{volume}{53},
  \bibinfo{number}{3} (\bibinfo{year}{1984}), \bibinfo{pages}{484--512}.
\newblock
\showISSN{0021-9991}
\href{https://doi.org/10.1016/0021-9991(84)90073-1}{doi:\nolinkurl{10.1016/0021-9991(84)90073-1}}


\bibitem[Birdsall and Fuss(1997)]%
        {BIRDSALL1997}
\bibfield{author}{\bibinfo{person}{Charles~K. Birdsall} {and}
  \bibinfo{person}{Dieter Fuss}.} \bibinfo{year}{1997}\natexlab{}.
\newblock \showarticletitle{Clouds-in-Clouds, Clouds-in-Cells Physics for
  Many-Body Plasma Simulation}.
\newblock \bibinfo{journal}{\emph{J. Comput. Phys.}} \bibinfo{volume}{135},
  \bibinfo{number}{2} (\bibinfo{year}{1997}), \bibinfo{pages}{141--148}.
\newblock
\showISSN{0021-9991}
\href{https://doi.org/10.1006/jcph.1997.5723}{doi:\nolinkurl{10.1006/jcph.1997.5723}}


\bibitem[Campbell and Neumann(2019)]%
        {Campbell_2019}
\bibfield{author}{\bibinfo{person}{John Campbell} {and} \bibinfo{person}{Tobias
  Neumann}.} \bibinfo{year}{2019}\natexlab{}.
\newblock \showarticletitle{Precision phenomenology with MCFM}.
\newblock \bibinfo{journal}{\emph{Journal of High Energy Physics}}
  \bibinfo{volume}{2019}, \bibinfo{number}{12} (\bibinfo{date}{Dec.}
  \bibinfo{year}{2019}).
\newblock
\showISSN{1029-8479}
\href{https://doi.org/10.1007/jhep12(2019)034}{doi:\nolinkurl{10.1007/jhep12(2019)034}}


\bibitem[Campbell et~al\mbox{.}(2011)]%
        {Campbell_2011}
\bibfield{author}{\bibinfo{person}{John~M. Campbell}, \bibinfo{person}{R.~Keith
  Ellis}, {and} \bibinfo{person}{Ciaran Williams}.}
  \bibinfo{year}{2011}\natexlab{}.
\newblock \showarticletitle{Vector boson pair production at the LHC}.
\newblock \bibinfo{journal}{\emph{Journal of High Energy Physics}}
  \bibinfo{volume}{2011}, \bibinfo{number}{7} (\bibinfo{date}{July}
  \bibinfo{year}{2011}).
\newblock
\showISSN{1029-8479}
\href{https://doi.org/10.1007/jhep07(2011)018}{doi:\nolinkurl{10.1007/jhep07(2011)018}}


\bibitem[Dearing et~al\mbox{.}(2024)]%
        {Dearing2024}
\bibfield{author}{\bibinfo{person}{Matthew~T. Dearing}, \bibinfo{person}{Yiheng
  Tao}, \bibinfo{person}{Xingfu Wu}, \bibinfo{person}{Zhiling Lan}, {and}
  \bibinfo{person}{Valerie Taylor}.} \bibinfo{year}{2024}\natexlab{}.
\newblock \bibinfo{title}{LASSI: An LLM-based Automated Self-Correcting
  Pipeline for Translating Parallel Scientific Codes}.
\newblock
\showeprint[arxiv]{2407.01638}~[cs.SE]
\urldef\tempurl%
\url{https://arxiv.org/abs/2407.01638}
\showURL{%
\tempurl}


\bibitem[Dhruv(2024)]%
        {akash_dhruv_2024_13879406}
\bibfield{author}{\bibinfo{person}{Akash Dhruv}.}
  \bibinfo{year}{2024}\natexlab{}.
\newblock \bibinfo{title}{{C}ode{S}cribe: 2024.09}.
\newblock
\href{https://doi.org/10.5281/zenodo.13879406}{doi:\nolinkurl{10.5281/zenodo.13879406}}


\bibitem[Dhruv and Dubey(2025)]%
        {Dhruv2025}
\bibfield{author}{\bibinfo{person}{Akash Dhruv} {and} \bibinfo{person}{Anshu
  Dubey}.} \bibinfo{year}{2025}\natexlab{}.
\newblock \showarticletitle{Leveraging Large Language Models for Code
  Translation and Software Development in Scientific Computing}. In
  \bibinfo{booktitle}{\emph{Proceedings of the Platform for Advanced Scientific
  Computing Conference}} (FHNW University of Applied Sciences and Arts
  Northwestern Switzerland, Brugg-Windisch, Switzerland)
  \emph{(\bibinfo{series}{PASC '25})}. \bibinfo{publisher}{Association for
  Computing Machinery}, \bibinfo{address}{New York, NY, USA},
  \bibinfo{pages}{1–9}.
\newblock
\showISBNx{9798400718861}
\href{https://doi.org/10.1145/3732775.3733572}{doi:\nolinkurl{10.1145/3732775.3733572}}


\bibitem[Dubey et~al\mbox{.}(2014)]%
        {dubey2014evolution}
\bibfield{author}{\bibinfo{person}{Anshu Dubey}, \bibinfo{person}{Katie
  Antypas}, \bibinfo{person}{Alan~C Calder}, \bibinfo{person}{Chris Daley},
  \bibinfo{person}{Bruce Fryxell}, \bibinfo{person}{J~Brad Gallagher},
  \bibinfo{person}{Donald~Q Lamb}, \bibinfo{person}{Dongwook Lee},
  \bibinfo{person}{Kevin Olson}, \bibinfo{person}{Lynn~B Reid},
  {et~al\mbox{.}}} \bibinfo{year}{2014}\natexlab{}.
\newblock \showarticletitle{Evolution of FLASH, a multi-physics scientific
  simulation code for high-performance computing}.
\newblock \bibinfo{journal}{\emph{The International journal of high performance
  computing applications}} \bibinfo{volume}{28}, \bibinfo{number}{2}
  (\bibinfo{year}{2014}), \bibinfo{pages}{225--237}.
\newblock


\bibitem[Dubey et~al\mbox{.}(2011)]%
        {DUBEY2011101}
\bibfield{author}{\bibinfo{person}{Anshu Dubey}, \bibinfo{person}{Katie
  Antypas}, {and} \bibinfo{person}{Christopher Daley}.}
  \bibinfo{year}{2011}\natexlab{}.
\newblock \showarticletitle{Parallel algorithms for moving Lagrangian data on
  block structured Eulerian meshes}.
\newblock \bibinfo{journal}{\emph{Parallel Comput.}} \bibinfo{volume}{37},
  \bibinfo{number}{2} (\bibinfo{year}{2011}), \bibinfo{pages}{101--113}.
\newblock
\showISSN{0167-8191}
\href{https://doi.org/10.1016/j.parco.2011.01.001}{doi:\nolinkurl{10.1016/j.parco.2011.01.001}}


\bibitem[Dubey et~al\mbox{.}(2009)]%
        {dubey2009extensible}
\bibfield{author}{\bibinfo{person}{Anshu Dubey}, \bibinfo{person}{Katie
  Antypas}, \bibinfo{person}{Murali~K Ganapathy}, \bibinfo{person}{Lynn~B
  Reid}, \bibinfo{person}{Katherine Riley}, \bibinfo{person}{Dan Sheeler},
  \bibinfo{person}{Andrew Siegel}, {and} \bibinfo{person}{Klaus Weide}.}
  \bibinfo{year}{2009}\natexlab{}.
\newblock \showarticletitle{Extensible component-based architecture for FLASH,
  a massively parallel, multiphysics simulation code}.
\newblock \bibinfo{journal}{\emph{Parallel Comput.}} \bibinfo{volume}{35},
  \bibinfo{number}{10-11} (\bibinfo{year}{2009}), \bibinfo{pages}{512--522}.
\newblock


\bibitem[Dubey et~al\mbox{.}(2012)]%
        {dubey2012imposing}
\bibfield{author}{\bibinfo{person}{Anshu Dubey}, \bibinfo{person}{Cristopher
  Daley}, \bibinfo{person}{J ZuHone}, \bibinfo{person}{Paul~M Ricker},
  \bibinfo{person}{Klaus Weide}, {and} \bibinfo{person}{Carlo Graziani}.}
  \bibinfo{year}{2012}\natexlab{}.
\newblock \showarticletitle{Imposing a Lagrangian particle framework on an
  Eulerian hydrodynamics infrastructure in FLASH}.
\newblock \bibinfo{journal}{\emph{The Astrophysical Journal Supplement Series}}
  \bibinfo{volume}{201}, \bibinfo{number}{2} (\bibinfo{year}{2012}),
  \bibinfo{pages}{27}.
\newblock


\bibitem[Github(2024)]%
        {copilot}
\bibfield{author}{\bibinfo{person}{Github}.} \bibinfo{year}{2024}\natexlab{}.
\newblock \bibinfo{title}{{Copilot}}.
\newblock
\urldef\tempurl%
\url{https://github.com/features/copilot}
\showURL{%
\tempurl}
\newblock
\shownote{Accessed: 2024-10-05}.


\bibitem[Godoy et~al\mbox{.}(2024)]%
        {godoy2024}
\bibfield{author}{\bibinfo{person}{William~F. Godoy}, \bibinfo{person}{Pedro
  Valero-Lara}, \bibinfo{person}{Keita Teranishi}, \bibinfo{person}{Prasanna
  Balaprakash}, {and} \bibinfo{person}{Jeffrey~S. Vetter}.}
  \bibinfo{year}{2024}\natexlab{}.
\newblock \showarticletitle{Large language model evaluation for
  high-performance computing software development}.
\newblock \bibinfo{journal}{\emph{Concurrency and Computation: Practice and
  Experience}} \bibinfo{volume}{36}, \bibinfo{number}{26}
  (\bibinfo{year}{2024}), \bibinfo{pages}{e8269}.
\newblock
\showeprint{https://onlinelibrary.wiley.com/doi/pdf/10.1002/cpe.8269}
\href{https://doi.org/10.1002/cpe.8269}{doi:\nolinkurl{10.1002/cpe.8269}}


\bibitem[OpenAI(2024)]%
        {chatgpt}
\bibfield{author}{\bibinfo{person}{OpenAI}.} \bibinfo{year}{2024}\natexlab{}.
\newblock \bibinfo{title}{{ChatGPT}}.
\newblock
\urldef\tempurl%
\url{https://chatgpt.com}
\showURL{%
\tempurl}
\newblock
\shownote{Accessed: 2024-10-05}.


\bibitem[Slack(2024)]%
        {slack}
\bibfield{author}{\bibinfo{person}{Slack}.} \bibinfo{year}{2024}\natexlab{}.
\newblock \bibinfo{title}{Balancing Old Tricks with New Feats: AI-Powered
  Conversion From Enzyme to React Testing Library at Slack}.
\newblock
\urldef\tempurl%
\url{https://slack.engineering/balancing-old-tricks-with-new-feats-ai-powered-conversion-from-enzyme-to-react-testing-library-at-slack/}
\showURL{%
\tempurl}
\newblock
\shownote{Accessed: 2024-10-05}.


\bibitem[Teranishi et~al\mbox{.}(2025)]%
        {teranishi2025}
\bibfield{author}{\bibinfo{person}{Keita Teranishi}, \bibinfo{person}{Harshitha
  Menon}, \bibinfo{person}{William~F. Godoy}, \bibinfo{person}{Prasanna
  Balaprakash}, \bibinfo{person}{David Bau}, \bibinfo{person}{Tal Ben-Nun},
  \bibinfo{person}{Abhinav Bhatele}, \bibinfo{person}{Franz Franchetti},
  \bibinfo{person}{Michael Franusich}, \bibinfo{person}{Todd Gamblin},
  \bibinfo{person}{Giorgis Georgakoudis}, \bibinfo{person}{Tom Goldstein},
  \bibinfo{person}{Arjun Guha}, \bibinfo{person}{Steven Hahn},
  \bibinfo{person}{Costin Iancu}, \bibinfo{person}{Zheming Jin},
  \bibinfo{person}{Terry Jones}, \bibinfo{person}{Tze~Meng Low},
  \bibinfo{person}{Het Mankad}, \bibinfo{person}{Narasinga~Rao Miniskar},
  \bibinfo{person}{Mohammad Alaul~Haque Monil}, \bibinfo{person}{Daniel
  Nichols}, \bibinfo{person}{Konstantinos Parasyris}, \bibinfo{person}{Swaroop
  Pophale}, \bibinfo{person}{Pedro Valero-Lara}, \bibinfo{person}{Jeffrey~S.
  Vetter}, \bibinfo{person}{Samuel Williams}, {and} \bibinfo{person}{Aaron
  Young}.} \bibinfo{year}{2025}\natexlab{}.
\newblock \bibinfo{title}{Leveraging AI for Productive and Trustworthy HPC
  Software: Challenges and Research Directions}.
\newblock
\showeprint[arxiv]{2505.08135}~[cs.SE]
\urldef\tempurl%
\url{https://arxiv.org/abs/2505.08135}
\showURL{%
\tempurl}


\bibitem[{Thomas Wolf, et al.}(2020)]%
        {huggingface}
\bibfield{author}{\bibinfo{person}{{Thomas Wolf, et al.}}}
  \bibinfo{year}{2020}\natexlab{}.
\newblock \bibinfo{title}{HuggingFace's Transformers: State-of-the-art Natural
  Language Processing}.
\newblock
\showeprint[arxiv]{1910.03771}~[cs.CL]
\urldef\tempurl%
\url{https://arxiv.org/abs/1910.03771}
\showURL{%
\tempurl}


\bibitem[Zi et~al\mbox{.}(2025)]%
        {zi2025}
\bibfield{author}{\bibinfo{person}{Yangtian Zi}, \bibinfo{person}{Harshitha
  Menon}, {and} \bibinfo{person}{Arjun Guha}.} \bibinfo{year}{2025}\natexlab{}.
\newblock \bibinfo{title}{More Than a Score: Probing the Impact of Prompt
  Specificity on LLM Code Generation}.
\newblock
\showeprint[arxiv]{2508.03678}~[cs.CL]
\urldef\tempurl%
\url{https://arxiv.org/abs/2508.03678}
\showURL{%
\tempurl}


\end{thebibliography}
\end{document}